# Review of MANETS Using Distributed Public-key Cryptography

Ms. Rajni[1] , Ms. Reena[2]

[1](M.Tech,Computer Science and Engineering Departmment,JMIT,Radaur/Kurukshetra University,India)
[2](Assistant Professor, Computer Science and Engineering Department, JMIT, Radaur/ Kurukshetra University, India)

**ABSTRACT :**

*Ensuring security is something that is not easily done as many of the demands of network security conflict with the demands of mobile networks, majorly because of the nature of the mobile devices (e.g. low power consumption, low processing load). The study of secure distributed key agreement has great theoretical and practical significance. Securing Mobile Ad-hoc Networks using Distributed Public-key Cryptography in pairing with Mobile Ad hoc Networks and various protocols are essential for secure communications in open and distributed environment.*

**KEYWORDS:**

*MANETS, Public key Cryptography, Key Management.Network Security,*

## 1. INTRODUCTION

In the last few years, the performances of wireless technologies have increased tremendously, thus opening new fields of application in the area of Computer Networks. One such fields concerns mobile ad hoc networks (MANETs). A mobile ad-hoc network (MANET) is a self-configuring network of mobile routers connected by wireless links—the union of which form an arbitrary topology. The wireless routers are in free random motion and they organize themselves arbitrarily. These types of networks operate in the absence of any fixed infrastructure. As the absence of any fixed infrastructures, it apparently becomes very difficult to make use of the existing routing techniques for network services, and poses various challenges in ensuring the security of the communication. As we know ensuring security is something that is not easily done as many of the demands of network security conflict with the demands of mobile networks, majorly because of the nature of the mobile devices (e.g. low power consumption, low processing load). The study of secure distributed key agreement has great theoretical and practical significance. Securing Mobile Ad-hoc Networks using Distributed Public-key Cryptography in pairing with Mobile Ad hoc Networks and various protocols are essential for secure communications in open and distributed environment.

### 1. 1 Introduction to MANETs

A MANET is a most promising and rapidly growing technology which is based on a self-organized and rapidly deployed network. Mobile Ad Hoc Networks (MANETS) are wireless mobile nodes that cooperatively form a network without infrastructure. In other words, ad hoc networking allows devices to create a network on demand without prior configuration [2, 3]. Thus, nodes within a MANET are involved in routing and forwarding information between neighbors, because there is no coordination or configuration prior to setup of a MANET. MANETs are self-configuring networks of mobile nodes without the presence of static infrastructure [4, 5]. They can also be diversified, which means that all nodes don't have the same capacity in term of resources (power consumptions, storage, computation, etc.). Due to its great features, MANET attracts different real world application areas where the networks topology changes very quickly.

Mobile Ad-hoc network is a set of wireless devices called wireless nodes, which dynamically link and transfer information. Wireless nodes can be personal computers (desktops/laptops) with wireless LAN cards, Personal Digital Assistants (PDA), or other types of wireless or mobile communication devices. In general, a wireless node can be any computing equipment that employs the air as the transmission medium. As shown, the wireless node may be physically attached to a person, a vehicle, or an airplane, to enable wireless communication among them.

### 1.2 Characteristics :

It is important to acknowledge the properties or characteristics of mobile ad hoc networks (MANETs).The characteristics [2] of MANETs and the possible applications are strongly related;

different applications demand MANETs with variants of the given characteristics.

- **Network Infrastructure:** There is no fixed or preexisting infrastructure in an ad hoc network: every network functions including routing, security, and network management are performed by the nodes themselves. Due to the nodes limited broadcasting range, data dissemination is achieved in a multi hop fashion; nodes can therefore be considered as hosts and routers. Node mobility and wireless connectivity allow nodes to spontaneously join and leave the network, which makes the network amorphous [2]. Security services must be able to scale seamlessly and remain available with changes in network topology.
- **Network Topology:** Nodes in ad hoc networks may be mobile resulting in a dynamic, weakly linked topology. Since node mobility is unrestricted, the topology may be unstable. The network demonstrates global mobility patterns, in any case, that may not be completely random.
- **Self-Organization:** MANETs cannot rely on any form of central administration or control; this is essential to avoid a single point of attack. A self-organized MANET cannot rely on any form of offline trusted third party (TTP); the network can thus be initialized by a distributed online TTP.
- **Limited Resources:** Nodes have limited computational, memory, and energy resources in contrast to their wired predecessors. Nodes are small hand-held devices (possibly "off-the-shelf" consumer electronics) that do not hinder user mobility. In an endeavor to keep the cost of these devices low, they are normally powered by a small CPU, accompanied by limited memory resources. As the devices are mobile, they are battery operated. This often results in short on times and the possibility of power failure due to battery weariness, perhaps during execution of a network-related function.
- **Poor Physical Security:** Nodes are mobile and therefore cannot be locked up in a secure room or closet. These small hand-held devices are easily compromised by either being lost or stolen. It is therefore highly probable than an adversary can physically compromise one or more nodes and perform any number of tests and analysis. The opposer can also use the nodes to attack distributed network services, such as a distributed online certificate authority [4]. Poor physical security is not as significant in "open" MANETs: the adversaries do not have to physically capture nodes to become an insider or to perform analysis on the protocols.

### 1.3 Current challenges in MANETS

In a mobile ad hoc network, all the nodes cooperate with each other to forward the coming packets in the network, and hence each node is effectively works as a router. Thus one of the most important issues is routing. This thesis focuses mainly on routing issues in ad hoc networks. In this section, some of the other issues in ad hoc networks are described :

Distributed network: A MANET is a distributed wireless network without any fixed infrastructure. That means no centralized server is required to maintain the state of the clients.

Dynamic topology: The nodes are mobile and hence the network is self-organizing. Because of this, the topology of the network keeps changing over time. Consequently, the routing protocols designed for such networks must also be adaptive to the topology changes.

Power awareness*:* Since the nodes in an ad hoc network typically run on batteries and are deployed in hostile terrains, they have stringent power requirements. This implies that the underlying protocols must be designed to conserve battery life.

Addressing scheme*:* The network topology keeps changing dynamically and hence the addressing scheme used is quite significant. A dynamic network topology requires a ubiquitous addressing scheme, which avoids any duplicate addresses. In wireless WAN environments, Mobile IP is being used. Because the static home agents and foreign agents are needed, hence, this solution is not suitable for ad hoc network.

Network size*:* The ability to enable commercial applications such as voice transmission in conference halls, meetings, etc., i the s an attractive feature of ad hoc networks. However, the delay involved in  underlying protocols places a strict upper bound on the size of the network.

Security: Security in an ad hoc network is extremely important in scenarios such as a battlefield . The five goals of security: availability, confidentiality, integrity, authenticity and non-repudiation are difficult to achieve in MANET, mainly because every node in the network participates equally in routing packets.

## 2. KEY MANAGEMENT IN MANETS

Key management can be defined as a set of techniques and procedures supporting the establishment and maintenance of keying relationships between authorized parties. In summary, key management integrates techniques and procedures to establish a service supporting various Initialization, Generation, distribution and updation of network keys.

The fundamental function of key management schemes is the establishment of keying material. Key establishment can be subdivided into key agreement and key transport. Key agreement allows two or more parties to derive shared keying material as a function of the information exchanged by and associated with, each of the protocol participants, such that no party can predetermine the resulting value. In key transport protocols, one party creates or otherwise obtains keying material, and securely transfers it to the other party or parties. Both key agreement and key transport can be achieved using either asymmetric techniques or symmetric. Hybrid key establishment scheme makes use of both symmetric and asymmetric techniques in an attempt to exploit the advantages of both techniques.

One of the major issues in such networks is its performance which arises in a dynamically changing topology; the mobile nodes are expected to be power-aware due to the bandwidth constrained network. Another major issue in such networks is security since every node participates in the operation of the network equally; malicious nodes are difficult to detect.

## 3.RELATED WORK

**[1]Samba Sessay, Zongkai Yang and Jianhua He,** This paper presents a coherent survey on ad hoc wireless networks, with the intent of serving as a quick reference to the current research issues in ad hoc networking. It starts with a background on the origin and development stages of ad hoc network, then summaries the features, capabilities, applications and design constraints of ad hoc network fully distinguishing it from traditional networks. The paper discuses a broad range of research issues such as Routing, Medium Access, Multicasting, Quality of service, TCP performance, Energy, Security and Bluetooth, outline the major challenges which have to be solved before widespread deployment of the technology is possible. Through this survey it would be seen that Ad hoc Networking presence an interesting research area inheriting the problems of wireless and mobile communications in their most difficult form.

**[2] Johann Van Der Merwe, J. Dawoudand Stephen McDonald,** Key management schemes based on the key predistribution techniques proposed for sensor networks may be another avenue to solve the key management problem in authority-based MANETs. Key management schemes are designed either for an "open" (self-organized) or "closed" (authority-based) network and consequently aimed at different applications. "Open" or fully self-organized MANETs have some inherent security implications and must be analyzed accordingly. It is therefore not always possible to compare schemes that assume the existence of a trusted authority with those that are fully self-organized. This study confirms that key management mechanisms proposed to guarantee the security of conventional networks are not necessarily suitable or adaptable to MANETs.

**[3] JeroenHoebeke, Ingrid Moerman, Bart Dhoedt and Piet Demeester** Current devices, their applications and protocols are solely focused on cellular or wireless local area networks (WLANs), not taking into account the great potential offered by mobile ad hoc networking. This type of network, operating as a stand-alone network or with one or multiple points of attachment to cellular networks or the Internet, coat the way for numerous new and exciting applications. Application scenarios involve, but are not limited to emergency and deliverance operations, conference or campus settings, car networks, personal networking, etc. This paper provides penetration into the potential applications of ad hoc networks and discusses the technological challenges that protocol designers and network developers are faced with.

**[4] Anne Marie Hegland, Eliwinjum, Stig F. Mjølsnes, ChunmingRong, Øivind Kure & Pal Spilling**
The wireless and dynamic nature of mobile ad hoc networks (MANETs) leaves them more vulnerable to security attacks than their wired similitude. The nodes act both as routers and as communication end points. This makes the network layer more predicate to security assail. A main contend is to judge whether or not a routing message originates from a trustworthy node. The clarification thus far is cryptographically signed messages. The general presumption is that nodes in possession of a valid secret key can be trusted. Consequently, a secure and effective key-management scheme is essential. Keys are also required for protection of application data. However, the direction here is on network-layer management information. This paper surveys the state of the art within key management for ad hoc networks, and examines their applicability for network-layer security. The analysis put some emphasis on their applicability in scenarios such as emergency and deliverance operations, as this work was initiated by a study of security in MANETs for emergency and rescue operations.

**[5]Eduardo da Silva, Aldri L. Dos Santos &Luiz Carlos P. Albini** Security is one of the major issues in MANETs. Their natural feature makes them vulnerable to numerous severe attacks. Several cryptographic mechanisms for MANETs can be found in the literature. Between them, identity-based cryptographic mechanisms and key management schemes are proposed to simplify key management and to reduce the memory storage cost. This paper presents the most important ID-based key management schemes, discussing their approaches, posture, and weaknesses, and comparing their main characteristics. It also presents the main ID based key management application fields on MANETs. In this way it can be utile for users and researchers as a starting point on ID-based key management and its possible uses in MANETs.

**[6] Kapil, Anil, and Sanjeev Rana. "Identity-Based Key Management in MANETs using Public Key Cryptography."** *International Journal of Security (IJS)* **3.1 (2009**): Wireless mobile Ad Hoc Networks (MANETs) are an emerging area of mobile computing. MANETs face serious security problems due to their unique characteristics such as mobility, dynamic topology and lack of central infrastructure support. In conventional networks, deploying a robust and reliable security scheme such as Public Key Infrastructure (PKI) requires a central authority or trusted third party to provide fundamental security services including digital certificates, authentication and encryption. In the proposed scheme, a secure identity-based key management scheme is proposed for networks in environments without any PKI. This scheme solved the security problem in the MANET and is also suitable for application to other wired network structures.

**[7]Zhang, Yanchao, et al. "AC-PKI: Anonymous and certificateless public-key infrastructure for mobile ad hoc networks."** *Communications, 2005. ICC 2005. 2005 IEEE International Conference on.* **Vol. 5. IEEE, 2005.** This paper studies public-key management, a fundamental problem in providing security support for mobile ad hoc networks. The infrastructureless nature and network dynamics of ad hoc networks make the conventional certificate based public-key solutions less suitable. To tackle this problem, a novel Anonymous and Certificateless Public-KeyInfrastructure (AC-PKI) for ad hoc networks was proposed. AC-PKI enables public-key services with certificateless public keys and thus avoids the complicated certificate management inevitable in conventional certificate-based solutions. To satisfy the demand for private keys during network operation, the secret-sharing technique to distribute a system master-key among a preselected set of nodes was employed, called D-PKGs, which offer a collaborative private-key-generation service. In addition, pinpoint attacks against D-PKGs detected.

## 4. CONCLUSION

The MANETs are the most promising and rapidly growing technology which is based on a self-organized and rapidly deployed network. Mobile Ad Hoc Networks (MANETS) are wireless mobile nodes that cooperatively form a network without infrastructure. In other words, ad hoc networking allows devices to create a network on demand without prior configuration. Thus, nodes within a MANET are involved in routing and forwarding information between neighbors, because there is no coordination or configuration prior to setup of a MANET. Due to Features provided by MANETS, MANET attracts different real world application areas where the networks topology changes very quickly.